\documentclass[journal,11pt,draftclsnofoot,onecolumn]{IEEEtran}

\usepackage{latexsym}
\usepackage{amsfonts}
\usepackage{graphicx} 
\usepackage{subfigure} 
\usepackage{amssymb}
\usepackage{stmaryrd}
\usepackage[numbers]{natbib}
\usepackage{enumerate}
\usepackage{amsmath}

\usepackage[usenames]{color}

\newcommand{\dfn}{\triangleq}

\newcommand{\ua}{\uparrow}
\newcommand{\nc}{\newcommand}
\nc{\da}{\downarrow} \nc{\hc}{\hat{c}} \nc{\hS}{\hat{S}}
\nc{\bra}{\langle} \nc{\ket}{\rangle} \nc{\eq}{equation (\ref}
\nc{\h}{\hat} \nc{\hT}{\h{T}}\nc{\be}{\begin{eqnarray}}
\nc{\ee}{\end{eqnarray}}\nc{\rd}{\textrm{d}}\nc{\e}{eqnarray}\nc{\hR}{\hat{R}}\nc{\Tr}{\mathrm{Tr}}
\nc{\tS}{\tilde{S}}\nc{\tr}{\mathrm{tr}}\nc{\8}{\infty}\nc{\lgs}{\bra\ua,\phi|}\nc{\rgs}{|\ua,\phi\ket}
\nc{\hU}{\hat{U}}\nc{\lfs}{\bra\phi|}\nc{\rfs}{|\phi\ket}\nc{\hZ}{\hat{Z}}\nc{\hd}{\hat{d}}\nc{\mD}{\mathcal{D}}
\nc{\bd}{\bar{d}}\nc{\bc}{\bar{c}}\nc{\mc}{\mathcal}\nc{\ea}{eqnarray}\nc{\mG}{\mathcal{G}}\nc{\bce}{\begin{center}}
\nc{\ece}{\end{center}}
\date{April 2013}

\begin{document}

\title{Extremely efficient generation of Gamma random variables for $\alpha \ge 1$}

\author{Luca Martino$^{1}$ and David Luengo$^{2}$\\
$^{1}$\textit{Dep. of Signal Theory and Communications, Universidad Carlos III de Madrid, Av. Universidad 30, 28911 Legan\'es, Spain}.\\
$^{2}$ \textit{Dep. Ing. de Circuitos y Sistemas, Univ. Polit\'ecnica de Madrid, Ctra. de Valencia km. 7, 28031 Madrid, Spain}.\\
 E-mail: luca@tsc.uc3m.es, luengod@ieee.org. \\
} 
\maketitle

\vspace{-1.5cm}
\begin{abstract}
The Gamma distribution is well-known and widely used in many signal processing and communications applications. In this letter, a simple and extremely efficient accept/reject algorithm is introduced for the generation of independent random variables from a Gamma distribution with any shape parameter $\alpha \geq 1$. The proposed method uses another Gamma distribution with integer $\alpha_p \le \alpha$, from which samples can be easily drawn, as proposal function. For this reason, the new technique attains a higher acceptance rate (AR) for $\alpha \geq 3$ than all the methods currently available in the literature, with $\textrm{AR} \to 1$ as $\alpha \to \infty$.
\end{abstract}

\section{Introduction}

The Gamma probability density function (PDF) is given by $p_o(x) = C_1 p(x)$, with $C_1 = \frac{\beta^{\alpha}}{\Gamma(\alpha)}$, and
\begin{equation}
	p(x) = x^{\alpha-1} \exp\left(-\beta x\right), \qquad x \ge 0,
\label{eq:GenGamma}
\end{equation}
where $\alpha > 0$ is the shape parameter, $\beta > 0$ is the rate parameter and $\Gamma(\alpha)$ is the Gamma function \citep[Chapter 4]{Dagpunar88}.
The Gamma distribution is well-known and widely used in different fields, such as Bayesian inference \citep{Griffin2010}, signal processing \citep{Jung2012} and digital communications \citep{Maaref2010}.
In particular, in communications it has been recently applied in the simulation of fading/shadowing channels using the Weibull-Gamma model \citep{Bithas2009,Anastasov2012} or the effects of the turbulent atmosphere in free-space optical links with the Gamma-Gamma approach, which requires two independent Gamma random variables \citep{Gappmair2007,Liu2010}.

All of the aforementioned applications need the generation of independent Gamma random variables (RVs), $X$, with arbitrary values of $\alpha$ and $\beta$, i.e., $X \sim G(\alpha,\beta)$.
When $\alpha$ is an integer, an exact sampler is available.
Indeed, if $\alpha = n \in \mathbb{N}^+$ the Gamma PDF becomes an {\it Erlang} PDF \citep{Dagpunar88}, and $X$ can be generated as the sum of $n = \alpha$ independent exponential RVs, i.e., $X=\sum_{i=1}^{n}E_i$, where each $E_i$ follows an exponential PDF with parameter $\beta$.
These exponentials can be easily obtained through the inversion method \citep[Chapter 3]{Dagpunar88}, allowing us to express $X$ as
\begin{equation}
	X = -\frac{1}{\beta}\sum_{i=1}^{\alpha}{\ln \left( U_i \right)} = -\frac{1}{\beta} \ln\left( \prod_{i=1}^{\alpha}U_i \right),
\label{Eq2}
\end{equation}
where the $U_i$ are uniform RVs, i.e., $U_i\sim \mathcal{U}([0,1])$.
For $\alpha \notin \mathbb{N}^+$, the problem of generating a Gamma RV $X$ is usually divided in two subcases: $\alpha<1$ and $\alpha\geq 1$.
Focusing on the second case, which is the one addressed in this paper, we note that an exact sampler does not exist, but several accept/reject methods have been introduced (see \citep{Dagpunar88,Devroye86,Dagpunar07} for a review of the approaches proposed).
%
%
Most of these methods consider only $\beta=1$, since, given $\widetilde{X} \sim G(\alpha,1)$, it can be easily shown that $X = \frac{1}{\beta} \widetilde{X} \sim G(\alpha,\beta)$.

In this letter, we develop an extremely efficient rejection sampler to draw independent samples from a Gamma PDF with $\alpha\geq 1$ and any value of $\beta$.
The proposed method outperforms all the alternative techniques reported in the literature in terms of acceptance rate (i.e., the key performance measure of a accept/reject methods) for $\alpha \geq 3$.
The main idea is using a suitable Gamma PDF with an integer $\alpha_p \le \alpha$ as a proposal density, from which samples can be easily drawn using Eq. \eqref{Eq2}.
Since the proposal is itself another Gamma PDF, it provides a very good fit of the target, thus attaining very high acceptance rates that tend to $100\%$ for $\alpha \rightarrow +\infty$, i.e., virtually providing exact sampling.

\vspace*{-3pt}

\section{Background: accept/reject algorithm} %

Rejection sampling (RS) is a classical technique for generating independent samples from an arbitrary target PDF, $p_o(x) = C_1 p(x)$ with $x\in \mathcal{D}$ and $C_1 = [\int_{\mathcal{D}} p(x)dx]^{-1}$, using an alternative simpler proposal PDF, $\pi_o(x) = C_2 \pi(x)$ with $x\in \mathcal{D}$ and $C_2 = [\int_{\mathcal{D}} \pi(x)dx]^{-1}$, such that $\pi(x)\geq p(x)$, i.e., $\pi(x)$ is a hat function w.r.t. $p(x)$.
RS works by generating samples from the proposal density, $x' \sim \pi_o(x)$, accepting them when $u' \le p(x')/\pi(x')$, with $u' \sim \mathcal{U}([0,1])$, and rejecting them otherwise.
The key performance measure for RS is the average acceptance rate (AR), $a_R = \int_{\mathcal{D}} \frac{p(x)}{\pi(x)} \pi_o(x) dx = \frac{C_2}{C_1} \le 1$.
The value of $a_R$ depends on how close the proposal is to the target, and determines the efficiency of the approach.
Hence, the main difficulty when designing an RS algorithm is finding a good hat function, $\pi(x) \geq p(x)$, such that $\pi(x)$ and $p(x)$ are as close as possible and drawing samples from $\pi_o(x)= C_2 \pi(x)$ can be done easily and efficiently.

\section{Novel technique}
In this letter, we consider as target density the PDF given in Eq. \eqref{eq:GenGamma} with $\alpha \geq 1$ and any $\beta > 0$.
As proposal PDF, we suggest using another Gamma density with different parameters, namely,
\begin{equation}
	\pi_o(x)\propto \pi(x) = K_p \mbox{ } x^{\alpha_p-1} \exp\left(-\beta_p x\right), \qquad x \ge 0,
\label{eq:proposal}
\end{equation}
where $\alpha_p =\lfloor \alpha \rfloor \le \alpha$, with $\lfloor \alpha \rfloor$ denoting the integer part of $\alpha\in [1,+\infty)$, and the remaining parameters ($\beta_p$ and $K_p$) adjusted to obtain the same location and value of the maximum for the proposal and the target.
On the one hand, for $\alpha \geq 2$,
\begin{align}
	\beta_p & = \frac{\alpha_p-1}{x_{\max}} = \beta \frac{\alpha_p-1}{\alpha-1},  \label{eq:Parprop2} \\
	K_p & = \frac{p(x_{\max})}{x^{\alpha_p-1} \exp\left(-\beta_p x\right)} \nonumber \\
		& = \exp\left(\alpha_p - \alpha\right)\left(\frac{\alpha - 1}{\beta}\right)^{(\alpha - \alpha_p)}, \label{eq:Parprop3}
\end{align}
where $x_{\max} = \frac{\alpha-1}{\beta}$ is the location of the single maximum of the Gamma PDF, obtained solving $\frac{dp(x)}{dx}=0$.
For $1\leq \alpha< 2$, $\alpha_p=1$, we set
\begin{align}
	\beta_p &= \frac{\beta}{\alpha},  \label{eq:Parprop4} \\
	K_p &= \exp(1-\alpha)\left(\frac{\alpha}{\beta}\right)^{\alpha-1}. \label{eq:Parprop5} 
\end{align}
In this case, they are obtained finding the exponential function tangent to $p(x)$ at the {\it optimal point}
$$
x^*=\frac{\alpha}{\beta}.
$$
This is an optimal value since it maximizes the acceptance rate, as shown in the Appendix \ref{BELLO_Sect}. When $\alpha=1$, the parameters are $\alpha_p=\alpha$, $\beta_p=\beta$ and $K_p=1$. In this case the unique intersection point between $\pi(x)$ and $p(x)$ is $x^*$. 
The values in Eqs. \eqref{eq:Parprop4}-\eqref{eq:Parprop5} can be easily obtained analytically computing the tangent straight line to the function $\log[p(x)]$ at $x^*$ (note that the Gamma pdf is a log-concave density).

\begin{figure}[!h]
\centerline{
\subfigure[]{\includegraphics[width=0.45\columnwidth]{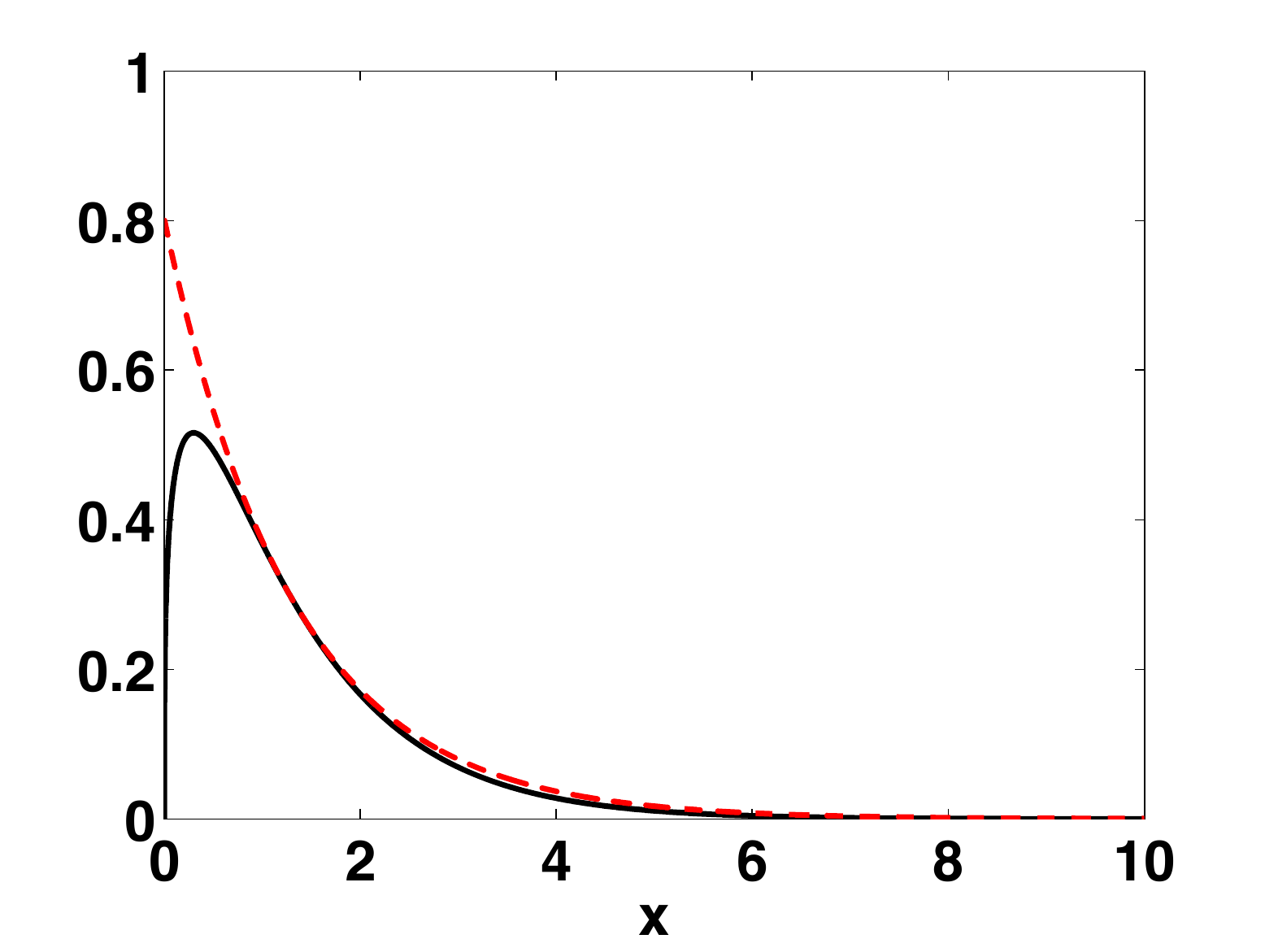}}
\subfigure[]{\includegraphics[width=0.45\columnwidth]{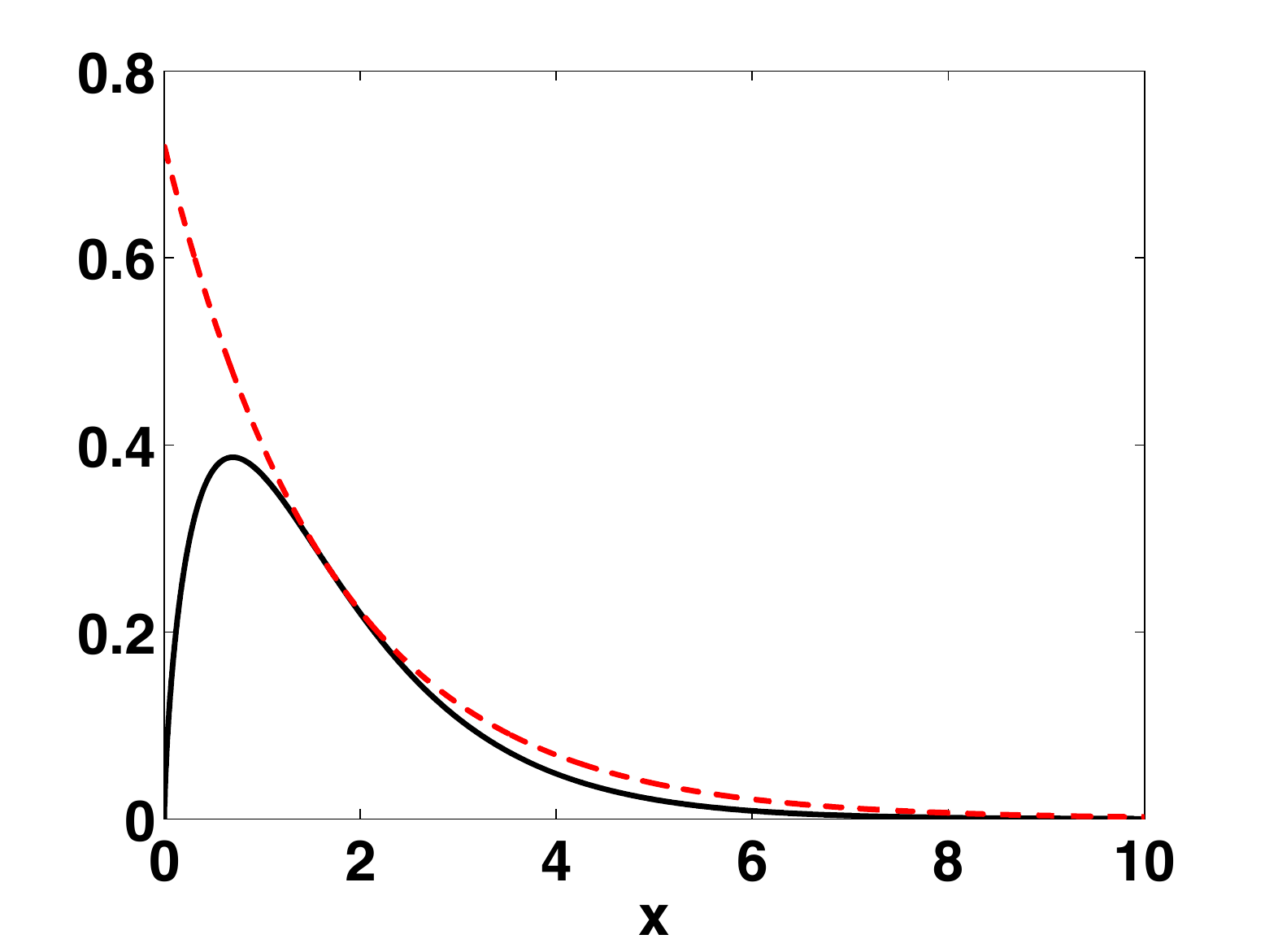}}
}
\centerline{
\subfigure[]{\includegraphics[width=0.45\columnwidth]{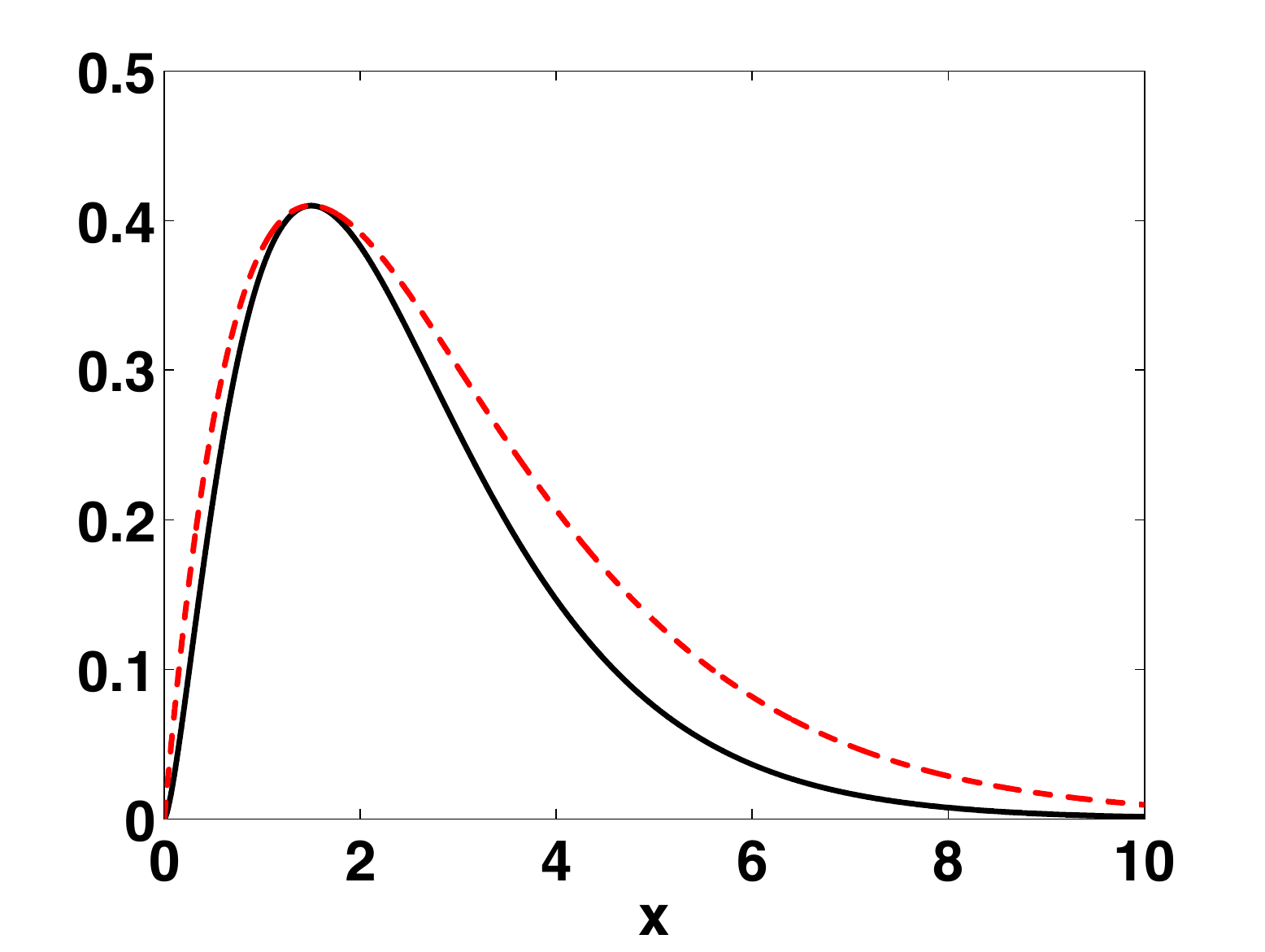}}
\subfigure[]{\includegraphics[width=0.45\columnwidth]{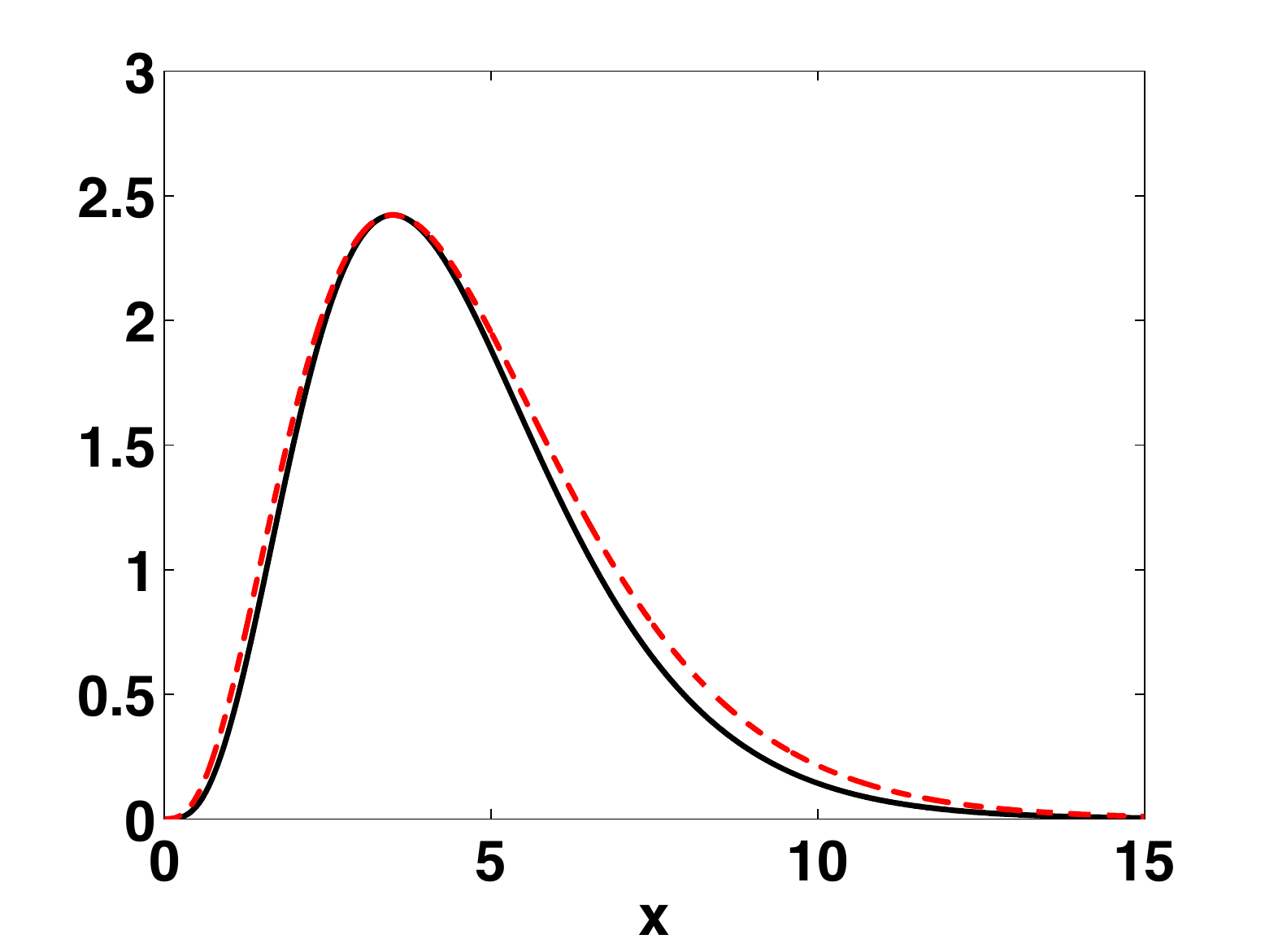}}
	}
	\caption{The target function $p(x)$ (solid line) and the proposed envelope function $\pi(x)$ (dashed line) with different values of the parameters: {\bf (a)} $\alpha=1.3$, $\beta=1$, {\bf (b)} $\alpha=1.7$, $\beta=1$, {\bf (c)} $\alpha=2.5$, $\beta=1$ and {\bf (d)} $\alpha=4.5$, $\beta=1$. }
\label{Figprop}
\end{figure}

Thanks to this choice of $\alpha_p$ and the parameters derived in Eqs.  \eqref{eq:Parprop2}-\eqref{eq:Parprop3}-\eqref{eq:Parprop4} and \eqref{eq:Parprop5}, we can ensure that: {\bf(a)} we can draw samples exactly from $\pi_o(x) \propto \pi(x)$ \citep{Dagpunar88}; {\bf(b)} $\pi(x) \geq p(x)$ for all $x \geq 0$, as proved in the following section. Figure \ref{Figprop} depicts some examples of envelope and target functions with different values of the parameters.

Our algorithm can be summarized in the following three steps: {\bf (1)} calculate the parameters of the proposal PDF, $\pi_o(x)\propto \pi(x)$; {\bf (2)} draw a sample $x'$ from $\pi_o(x)$ using the direct approach described in Eq. \eqref{Eq2}, i.e., generate $\alpha_p$ independent uniform RVs, $u_i \sim \mathcal{U}([0,1])$ with $i=1,...,\alpha_p$, and set
\begin{equation}
	x'=-\frac{1}{\beta_p}\ln \left( \prod_{i=1}^{\alpha_p} u_i \right);
\end{equation}
{\bf (3)} accept $x'$ with probability $p(x')/\pi(x')$, discarding it otherwise.
Repeat steps (2) and (3) until the desired number of samples have been obtained.

\section{Proof of the RS inequality}
\subsection{Case $\alpha\geq  2$}
Consider  first the proposal pdf for $\alpha\geq 2$.In order to apply the RS technique, we need to ensure that $\pi(x) \geq p(x)$, i.e.,
\begin{equation}
	K_p x^{\alpha_p-1} \exp\left(- \beta_p x\right)\geq x^{\alpha-1} \exp\left(-\beta x\right),  \qquad \forall x \ge 0.
\label{Cazzo1}
\end{equation}
For $x > 0$, Eq. \eqref{Cazzo1} can be rewritten alternatively as
\begin{equation}
	K_p \exp\left(\Omega x\right)\geq x^{\alpha-\alpha_p},
\label{Cazzo2}
\end{equation}
where $\Omega\dfn \beta-\beta_p$ and $x^{\alpha-\alpha_p}$ presents a sub-linear growth, since $0 \le \alpha-\alpha_p < 1$.
Finally, taking the logarithm on both sides of \eqref{Cazzo2},
\begin{equation}
	\ln(K_p) + \Omega x \geq (\alpha-\alpha_p) \ln(x).
\label{Cazzo3}
\end{equation}
Now, since $\alpha \ge \alpha_p$ and $\beta_p$ is given by \eqref{eq:Parprop2}, we note that
\begin{equation}
	\Omega = \beta-\beta_p = \beta\left(1-\frac{\alpha_p-1}{\alpha-1}\right) \geq 0.
\label{Cazzo4}
\end{equation}
Hence, the linear function on the left hand side of \eqref{Cazzo3} is increasing.
Moreover, since $\alpha \ge \alpha_p$, the logarithmic function on the right hand side of \eqref{Cazzo3} is increasing and .
Consequently, since both functions are increasing and concave for $x > 0$ (i.e., their second derivatives are lower or equal than zero), they can have at most two intersection points.
Indeed, in the sequel we show that they are tangent at $x = x_{\max}$, which is the only contact point between both curves for $x > 0$.
In order to prove this, we need to show: {\bf(1)} that both functions are equal at $x=x_{\max}$, i.e.,
\begin{equation}
	\ln(K_p) + \Omega x_{\max} = (\alpha-\alpha_p) \ln(x_{\max}) = (\alpha-\alpha_p) \ln\left(\frac{\alpha-1}{\beta}\right),
\end{equation}
which is fulfilled by construction of the proposal, as $K_p$ and $\beta_p$ are set to achieve $\pi(x_{\max}) = p(x_{\max})$;
{\bf(2)} that their first derivatives are equal, i.e.,
\begin{equation}
	\frac{d (\ln(K_p) + \Omega x)}{dx}\Bigg|_{x=x_{\max}} = \frac{d ((\alpha-\alpha_p) \ln x)}{dx}\Bigg|_{x=x_{\max}},
\end{equation}
and taking the derivatives we obtain
\begin{equation}
 \Omega = \frac{\alpha-\alpha_p}{x_{\max}} = \frac{\beta(\alpha-\alpha_p)}{\alpha-1} = \beta\left(1-\frac{\alpha_p-1}{\alpha-1}\right),
\end{equation}
which is the result given by Eq. \eqref{Cazzo4}.

Consequently, since $x$ grows faster than $\ln(x)$, we can guarantee that $\ln(K_p) + \Omega x \ge (\alpha-\alpha_p) \ln(x)$ for $x > 0$, with equality only at $x = x_{\max}$.
Hence, the RS inequality in \eqref{Cazzo1} is satisfied and the proposal is indeed a hat function for $p(x)$, i.e., $\pi(x) \geq p(x)$, with equality only at $x = 0$,
$x = x_{\max}$ and $x \to +\infty$.

\subsection{Case $1 \leq \alpha< 2$}
For, $1 \leq \alpha< 2$, we have $\alpha_p=1$ and the proposal pdf is built computing the tangent straight line $r(x)=-\beta_p x+\log[K_p]$ tangent to $\log[p(x)]$ at a generic point $x^*$ (for the optimal choice of this point, see the Appendix \ref{BELLO_Sect}). Then, the envelope function  
is 
$$
\pi(x)=\exp(r(x))=K_p \exp(-\beta_p x).
$$
For the log-concavity of $p(x)$, i.e.,
$$
\frac{d^2\log[p(x)]}{dx^2}\geq 0,
$$
we have 
$$
r(x)\geq \log[p(x)], 
$$
and since the exponential function is a monotonically increasing transformation, we have $\pi(x) \geq p(x)$, that is the needed condition to apply the RS technique.

\section{Acceptance rate of the novel RS scheme}
The acceptance rate of the novel scheme can be calculated analytically. Indeed, in general, we have 
$$
a_{R}=\frac{\int_{0}^{+\infty} p(x)dx}{\int_{0}^{+\infty} \pi(x)dx},
$$
so that 
\begin{gather}
\label{CasiIncreible}
a_{R}=\left\{ 
\begin{split}
 \frac{\Gamma(\alpha)}{ \alpha^{\alpha}e^{1-\alpha}}, \quad\quad\quad\quad\quad & \mbox{ for } 1 \leq \alpha <2,  \\
\left(\frac{e}{\alpha - 1}\right)^{(\alpha - \alpha_p)}\frac{\Gamma(\alpha)}{\Gamma(\alpha_p)}, \quad& \mbox{ for }  \alpha \geq 2,
\end{split}
\right.
\end{gather}
that is independent from the parameter $\beta$.

\section{Results}

\begin{figure}[!h]
\centerline{
	\includegraphics[width=0.7\columnwidth]{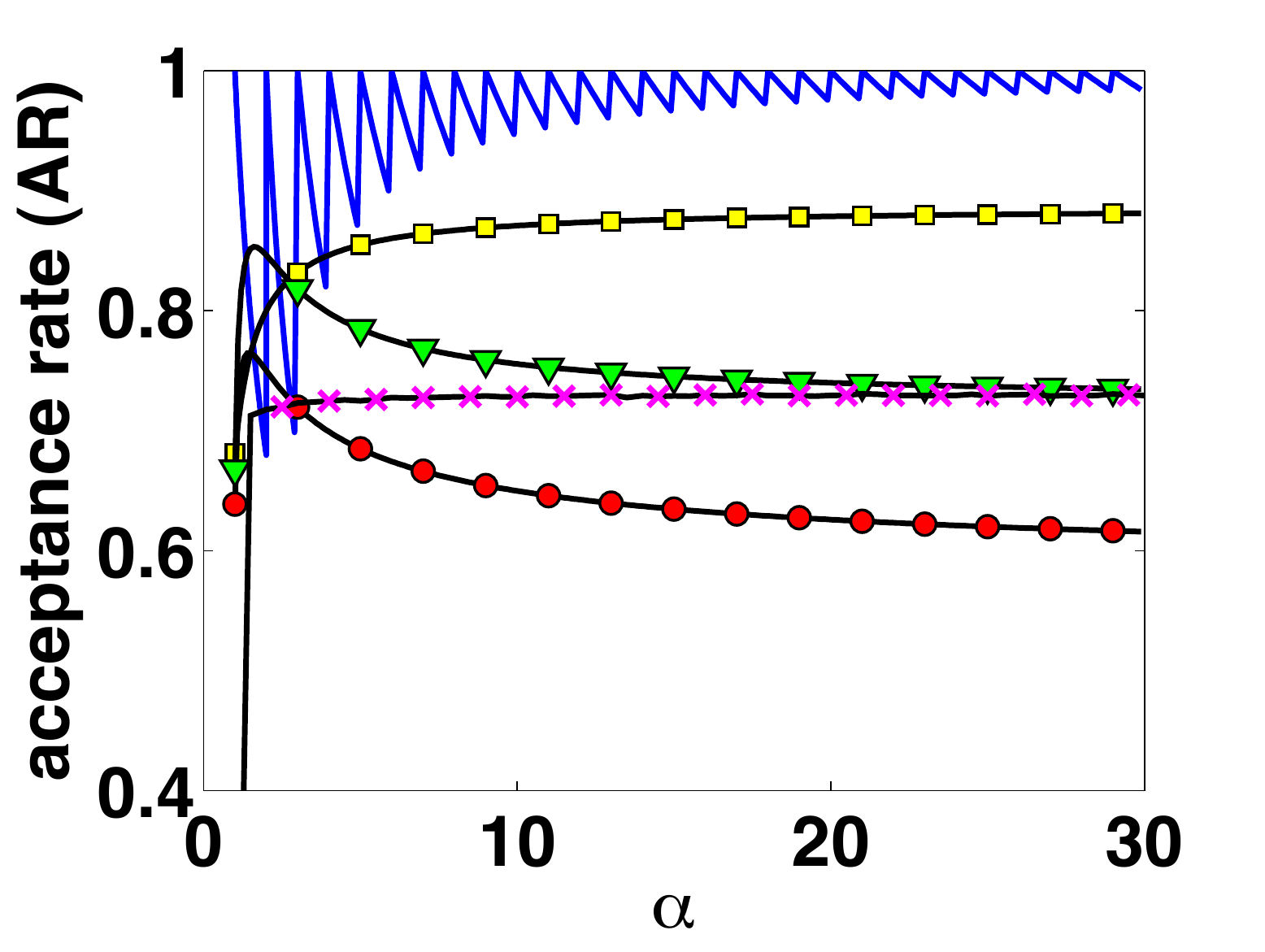}
	}
	\caption{Acceptance rate (AR) using our method M1 (continuous line), M2 (squares), M3 (circles), M4 (triangles) and M5 (x-marks), for $1 \le \alpha \le 30$ and $\beta=1$.}
\label{FigSimu}
\end{figure}

Gamma generators in the literature are usually designed and compared for $\beta=1$, without loss of generality.
%
%
Hence, in order to obtain a fair comparison we only consider $\beta=1$, although our approach is valid for any value of $\beta$.
We have compared the acceptance rate (AR), $a_R$, of the different algorithms described below \citep{Dagpunar88,Devroye86,Dagpunar07}:

\begin{itemize}

\item {\bf Our method (M1):}  For $1\leq \alpha <2$, the AR is indicated as $a_{R1}$ and is given in Eq. \eqref{CasiIncreible}. Note that, for $\alpha \rightarrow +\infty$, we obtain $a_{R1}\rightarrow 1$, i.e., our approach provides exact sampling asymptotically.

\item {\bf Log-logistic method~~(M2) \citep{Cheng77}:} The proposal PDF is $\pi(x)=K_1 \frac{x^{\lambda-1}}{(\mu+x^\lambda)}$, with $\lambda=\sqrt{2\alpha-1}$, $\mu=\alpha^\lambda$ and $K_1=4\alpha^{(\alpha+\lambda)}e^{-\alpha}$. The theoretical AR is $a_{R2}=\frac{\Gamma(\alpha)\mu\lambda}{K_1}$. For $\alpha \rightarrow +\infty$, $a_{R2}\rightarrow \frac{\sqrt{\pi}}{2}\approx 0.88$. 

\item {\bf Cauchy method (M3) \citep{Ahrens74}:} The proposal is $\pi(x)=K_2\frac{\lambda}{\lambda^2+(x-a)^2}$, with $\lambda=\sqrt{2\alpha-1}$, $a=\alpha-1$ and $K_2=\frac{c}{2\pi} \lambda(a-1)^{(a-1)}e^{1-a}$, where $c=\pi+2\arctan(a/\lambda)$. The AR is $a_{R3}=\frac{\Gamma(\alpha)}{\pi K_2}$ and we have $a_{R3}\rightarrow \frac{1}{\sqrt{\pi}}\approx 0.56$, for $\alpha \rightarrow +\infty$.

\item {\bf T-student method (M4) \citep{Best78}:} The proposal is $\pi(x)=K_3(1+0.5(\frac{x-\alpha+1}{\eta})^2)^{-3/2}$ with $\eta=\sqrt{\frac{3\alpha-0.75}{2}}$ and $K_3= (\alpha-1)^{\alpha-1}e^{1-\alpha}$. It can be shown that $a_{R4}\rightarrow \sqrt{\frac{\pi}{6}}\approx 0.72$ for $\alpha \rightarrow +\infty$.

\item {\bf Modified Ratio-of-Uniforms (RoU) (M5) \citep{Kinderman80}:} It is a variant of the RoU scheme \citep{Dagpunar88,Devroye86},  relocating the mode of $p(x)$ at $x=0$. The asymptotic AR is $a_{R4}\rightarrow \frac{\sqrt{e\pi}}{4}\approx 0.73$, for $\alpha \rightarrow +\infty$. The AR, $a_{R4}$, is almost constant for all values of $\alpha > 1$, as shown in Figure \ref{FigSimu}.  

\end{itemize}

Figure \ref{FigSimu} shows the ARs of all the techniques described above, obtained empirically after drawing $N = 6 \cdot 10^5$ independent samples, for different values of $\alpha$.
For $1\le \alpha \le 1.35$ our the technique M1 provide the best results whereas for $1.35\le \alpha \le 2$ the best technique is M4.
%
For $2 \le \alpha \le 2.37$ the newly proposed approach (M1) provides the highest AR  while for $2.37 < \alpha < 3$, the best method is M2 or M4 depending of the value of $\alpha$.
For $\alpha \ge 3$, our technique (M1) is extremely efficient, outperforming (expect for $3.87\leq \alpha<4$ where M2 is slightly better) the rest of the methods and providing the best results ever reported in the literature. The minimum AR obtained with M1 is $\approx 0.68$ for $\alpha\approx 1.99$.
Furthermore, our technique provides exact sampling (i.e., $a_{R1} \to 1$) asymptotically as $\alpha \rightarrow +\infty$.

%

\section{Conclusions}

We have developed a rejection sampling (RS) scheme for generating Gamma random variables, with arbitrary values of $\alpha\geq 1$  and $\beta$, where the proposal PDF is itself another Gamma density.
The proposed algorithm is simple and extremely efficient, providing the best acceptance rates ever reported in the literature for $\alpha\geq 3$.
%


\section{Acknowledgement}
This work has been partly financed by the Spanish government, through the CONSOLIDER-INGENIO 2010 program (CSD2008-00010), as well as projects COMPREHENSION (TEC2012-38883-C02-01) and DISSECT (TEC2012-38058-C03-01).

\appendix
\subsection{Optimal choice of the  tangent point for $1\leq \alpha <2$ }
\label{BELLO_Sect}
First, we recall the notation $p_o(x) = C_1 p(x)$ with 
$$C_1= \left[\int_{\mathcal{D}} p(x)dx\right]^{-1},$$
and $\pi_o(x) = C_2 \pi(x)$ with
$$C_2=\frac{1}{I_\pi}= \left[\int_{\mathcal{D}} \pi(x)dx\right]^{-1}.$$
The acceptance rate (AR) is
\begin{gather}
\begin{split}
a_R &= \int_{\mathcal{D}} \frac{p(x)}{\pi(x)} \pi_o(x) dx  \\
&=\frac{\int_{\mathcal{D}} p(x)dx}{\int_{\mathcal{D}} \pi(x)dx}=\frac{C_2}{C_1}=\frac{1}{C_1I_\pi} \le 1.
\end{split}
\end{gather}
Since the are below the target $p(x)$ is given (then fixed), the only way to increase the AR is diminishing the area $I_\pi$ below the envelope function, i.e., we desire to build a function $\pi(x)$ such that satisfies jointly both conditions
\begin{gather}
\left\{
\begin{split}
&I_\pi^*=\min_{\pi(x)} \int_{\mathcal{D}} \pi(x)dx, \\
&\pi(x)\geq p(x).
\end{split}
\right.
\end{gather}
For $1\leq \alpha <2$, the novel technique uses as proposal function of the form in Eq. \eqref{eq:proposal} with $\alpha_p=1$, and the other parameters, as function of a generic tangent point $\theta$, are
$$
\beta_p(\theta)=-\frac{\alpha- 1}{\theta} + \beta, \quad \quad \theta\neq 0
$$
$$
K_p(\theta)=\theta^{\alpha-1}e^{1-\alpha}.
$$
Therefore, our proposal for $1\leq \alpha <2$  has the following form 
$$
\pi(x)=K_p(\theta) \exp\left(-\beta_p(\theta) x\right), \quad x\in \mathcal{D}=\mathbb{R}^+, \quad \theta\in \mathbb{R}^+ \setminus \{0\},
$$
then
\begin{gather}
\begin{split}
I_\pi(\theta)&=\int_{0}^{+\infty} K_p(\theta) \exp\left(-\beta_p(\theta) x\right) dx= \frac{K_p(\theta)}{ \beta_p(\theta)} \\
&=\frac{\theta^{\alpha-1}e^{1 - \alpha}}{-\frac{\alpha- 1}{\theta} + \beta},  \quad   \theta\neq 0.
\end{split}
\end{gather}
The value of $\theta$ that minimizes $I_\pi(\theta)$ is a solution of the equation 
\begin{gather}
\begin{split}
&\frac{dI_\pi(\theta)}{d\theta}=0,\\
&\frac{\theta^{\alpha-1}(\alpha - \beta\theta)(1-\alpha)e^{1 - \alpha}}{(\beta\theta - \alpha + 1)^2}=0.
\end{split}
\end{gather}
 The solution are $\theta^*=0$ (that is not admissible) and
$$
x^*=\theta^*=\frac{\alpha}{\beta}.
$$
Choosing this value $x^*$ as tangent point to construct the envelope function, 
$$
\pi(x)=\left(\frac{\alpha}{\beta}\right)^{\alpha-1}e^{1-\alpha} \exp\left(- \frac{\beta}{\alpha} x\right),
$$ 
we maxime the AR of the RS scheme. 
Since $C_1=\frac{\beta^\alpha}{\Gamma(\alpha)}$, the AR in this case is
\begin{gather}
\begin{split}
a_R&=\frac{\Gamma(\alpha)\beta_p(x^*)}{\beta^\alpha K_p(x^*)} \\
&= \frac{\Gamma(\alpha)\frac{\beta}{\alpha}}{\beta^\alpha \left(\frac{\alpha}{\beta}\right)^{\alpha-1}e^{1-\alpha}}=\frac{\Gamma(\alpha)}{\beta^\alpha \left(\frac{\alpha}{\beta}\right)^{\alpha}e^{1-\alpha}} \\
&=\frac{\Gamma(\alpha)}{ \alpha^{\alpha}e^{1-\alpha}},
\end{split}
\end{gather}
independent from the parameter $\beta$.
 

\bibliographystyle{plain}
\bibliography{bibliografia,biblioFading}
\end{document}